\documentclass[preprint]{aastex}

\shorttitle{The Geminga Pulsar}
\shortauthors{Alexander ~A. ~Ershov}

\begin{document}

\title{
On radio emission of the Geminga pulsar and RBS~1223 at the
frequency of 111~MHz }

\author{Alexander A. Ershov \altaffilmark{1,2}}

\altaffiltext{1}{Pushchino Radio Astronomy Observatory, Astro Space
Center, Lebedev Physical Institute, Russia}

\altaffiltext{2}{E-mail:  ershov@prao.ru}

\begin{abstract}
I have searched for pulsed radio emission from the Geminga pulsar
and for the nearby isolated neutron star 1RX~J1308.6+2127
(RBS~1223) at the frequency of 111~MHz. No pulsed signals were
detected from these sources. Upper limits for mean flux density
are 0.4~-~4~mJy for the Geminga pulsar and 1.5~-~15~mJy for
RBS~1223 depending on assumed duty cycle (.05~-~.5) of the
pulsars.
\end{abstract}


\section{Introduction}

The gamma pulsar Geminga was discovered in 1972 with the help of
SAS-2 satellite (\cite{fic75}). It is the second brightest
gamma-source on the sky at energies over 100~MeV and it was
investigated in all bands of electro-magnetic spectrum. Its
identification as a pulsar was not secure until the detection of
X-ray (\cite{hal92}) and gamma-ray pulsations (\cite{ber92}). In
1990-s three groups from Pushchino Radio Astronomy Observatory
reported on discovery of pulsed radio emission from the Geminga
pulsar at 102.5~MHz (\cite{kuz97}, \cite{mal97}, \cite{shi98})
with flux density from 30~mJy to 100~mJy and dispersion measure
about of 3~$\rm pc~cm^{-3}$. But numerous searches for pulsed
radio emission from the Geminga pulsar at higher frequencies had
no positive results (see \cite{kas99} and references therein). I
searched for pulsed radio emission from this pulsar at the
frequency of 111~MHz.

1RX~J1308.6+2127 (also known as RBS~1223) is a nearby isolated
neutron star (\cite{sch99}) with the period of 10.3~s
(\cite{hab04}), identified also as a very faint optical object
(\cite{kap02}). Recently \cite{mal05} reported on discovery of
pulsed radio emission from this neutron star at a frequency of
111~MHz with a mean flux density about of 50~mJy. I searched for
pulsed radio emission from this neutron star as well: at the same
frequency and with the same radio telescope.

\section{Observations and Data Reduction}

The observations were performed from November 1999 through March
2007 with the Large Phase Array (BSA) radio telescope at Pushchino
Radio Astronomy Observatory with an effective area at zenith of
about 15,000 square meters. One linear polarization was received.
I used 128-channel receiver with a channel bandwidth of 20~kHz and
a center frequency of 110.59~MHz. The observations were carried
out in the mode of recording individual pulses. The sampling
interval was 2.56~ms at the receiver time constant $\tau = 3$~ms
for the Geminga pulsar and 5.0~ms at the receiver time constant
$\tau = 10$~ms for the RBS~1223. Since BSA radio telescope is a
transit one, the duration of one observing session is limited to
$3.2~/~cos(\delta)$~min. A total of 600 observational sessions was
carried out for each pulsar. These observations contain 441,000
pulsar periods for the Geminga pulsar and 12,000 pulsar periods
for RBS~1223. Since the middle of 2004 antenna was calibrated by
observations of the 3C~452 source, flux density of which at
111~MHz is considered as 91~Jy. The nearby (on the sky) pulsar PSR
B0626+24 was observed as a test pulsar for the Geminga pulsar
observations.

At primary processing of day observation session, a mean value was
deduced from time-series in each frequency channel and a result
was divided by mean square deviation of the channel. Then records
were reviewed to reveal interference, namely: records of all
channels were averaged without compensation for dispersion delay
(as ground interferences have no dispersion delay) and if
interference was revealed (with signal to noise ratio of seven or
higher) then corresponding values were substituted with zero at
all channels. Further: folding, that is summation of periods in
record of each channel, was performed; at that, period meaning for
a special day of observations was calculated on the base of recent
ephemeris (\cite{jac05}, \cite{hal07} for the Geminga pulsar and
\cite{kap05} for RBS~1223). And finally: compensation of
dispersion delay was performed for each channel; at that,
dispersion measure was searched in the range from 0 to 40~$\rm
pc~cm^{-3}$ with spacing of $1~\rm pc~cm^{-3}$. The proposed
distance (160~pc) to the Geminga pulsar corresponds to dispersion
measure about of $DM~=~3~\rm pc~cm^{-3}$. No statistically
(signal-to-noise ratio $S/N~>~5$) meaning radio emission was found
in any series of observations for both pulsars.

\section{Results}

\subsection{The Geminga Pulsar}

To improve the sensitivity of the search, all 600 observational
sessions were averaged together by time reference in accordance
with updated ephemeris (\cite{jac05}, \cite{hal07}) and with above
mentioned searches of dispersion measure. I did not reveal any
significant radio emission at this processing either. Examples of
resulting average (for all 600 sessions) pulse profiles for a
number of dispersion measure values are presented at the Fig~1.
Value of upper limit ($5 \sigma$, at smoothing to the time
resolution of 10~ms) for a peak flux density equals to 8~mJy. A
correspondent value of the mean (by period) upper limit is within
0.4 to 4 mJy range depending on an assumed (.05 to .5) pulse duty
cycle.

Fig~2 shows profile of the Geminga pulsar (for all 600 series;
smoothed by 4 points and 2 periods are presented) for dispersion
measure of 3~$\rm pc~cm^{-3}$ together with average profile of the
test pulsar PSR B0626+24. The mean flux density of PSR B0626+24 is
60~mJy, that is approximately the same as was declared for the
Geminga pulsar. Unfortunately, my results do not confirm results
of my colleagues.

\subsection{1RX~J1308.6+2127 (RBS~1223)}

To improve the sensitivity of the search, all 600 observational
sessions were averaged together by time reference in accordance
with ephemeris from paper of \cite{kap05} and with above mentioned
searches of dispersion measure. I did not reveal any significant
radio emission at this processing either. Examples of resulting
means (for all 600 sessions) of pulse profiles for a number of
dispersion measure values are presented at the Fig~3. Value of
upper limit ($5 \sigma$) for a peak flux density equals to 30~mJy
(at smoothing to the time resolution of 20 ms). A correspondent
value of the mean (by period) upper limit is within 1.5 to 15~mJy
range depending on a proposed (.05 to .5 period) pulse duration.
Unfortunately, my results do not confirm results of
Malofeev~et~al.~(2005).

\section{Conclusions}

The search of pulsed radio emission from the Geminga pulsar and
1RX~J1308.6+2127 (RBS~1223) at the frequency of 111~MHz give no
positive results. Upper limits for mean flux density are
0.4~-~4~mJy for the Geminga pulsar and 1.5~-~15~mJy for RBS~1223
depending on assumed duty cycle (.05 - .5) of the pulsar.

\section{Acknowledgements}

I am grateful to the staff of the Pushchino Radio Astronomy
Observatory for help in preparation and performing of
observations. I am grateful to Prof. Jules ~P. ~Halpern (Columbia
University) for sending me updated ephemeris of the Geminga pulsar
before publication. This paper was submitted as a poster for the
40 Years of Pulsars Conference. But because of non-arrival of the
author the conference organizers would not let another person to
mount this poster, unfortunately.


\begin{figure}
\center
\includegraphics[scale=1.00]{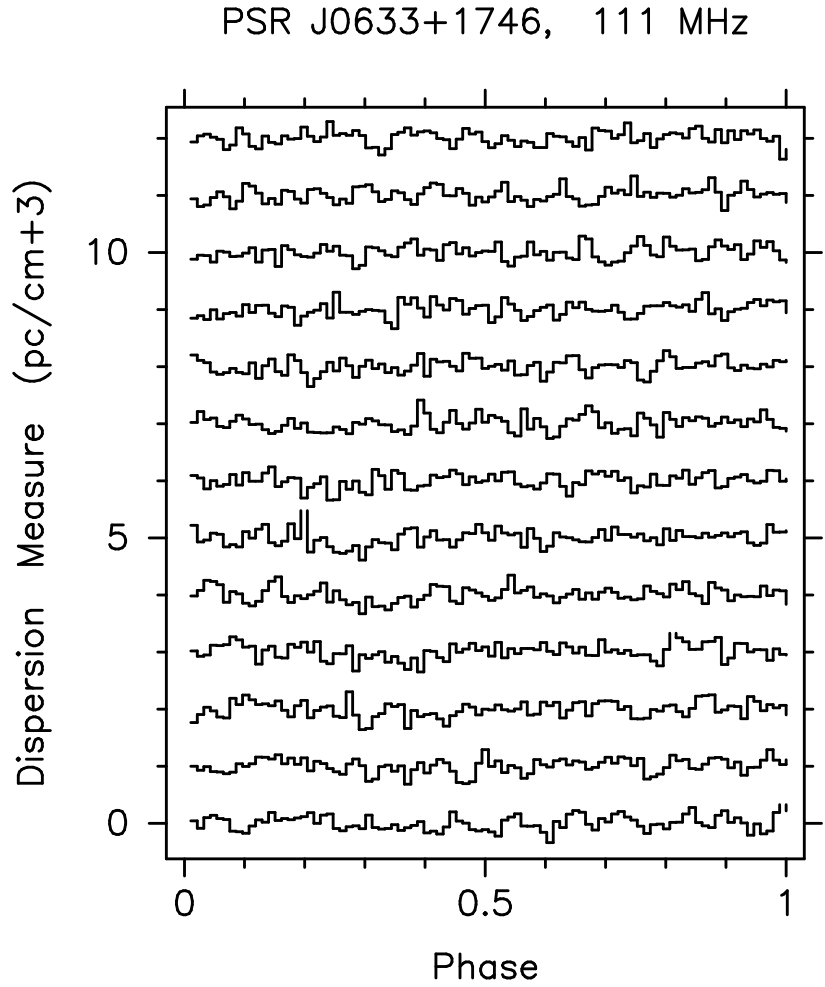}
\caption{
The average profiles of the Geminga pulsar at the
frequency of 111~MHz for a number of dispersion measure values.
\label{fig1}}
\end{figure}

\begin{figure}
\center
\includegraphics[scale=1.00]{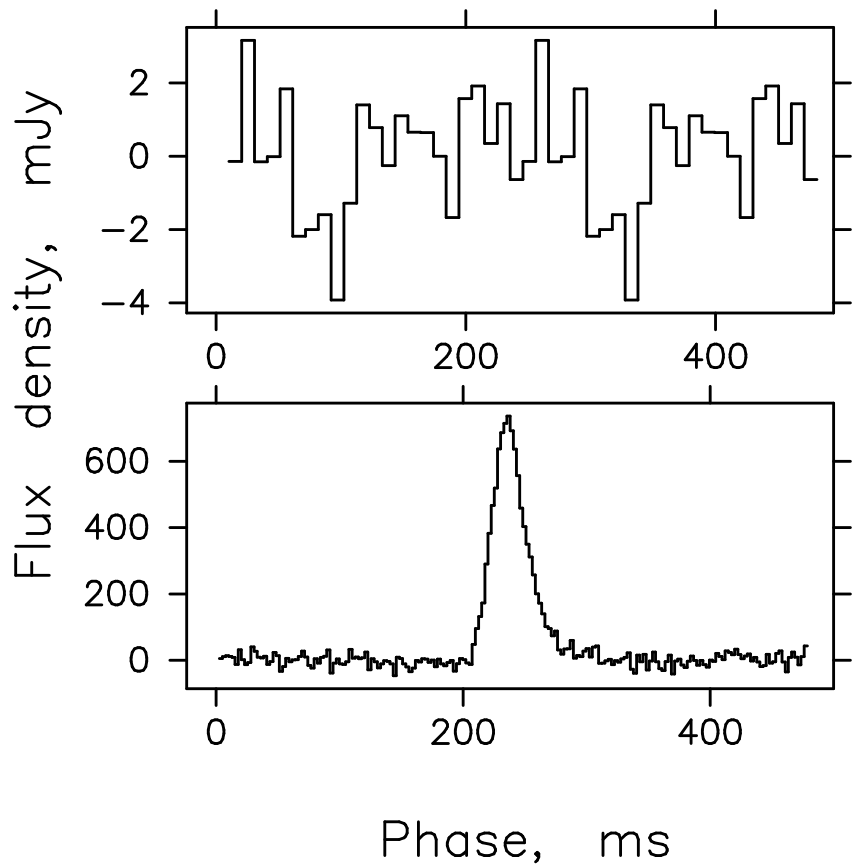}
\caption{
The average profile of the Geminga pulsar for dispersion
measure of 3~$\rm pc~cm^{-3}$ at the frequency of 111~MHz together
with the profile of test pulsar PSR B0626+24.
\label{fig2}}
\end{figure}

\begin{figure}
\center
\includegraphics[scale=1.20]{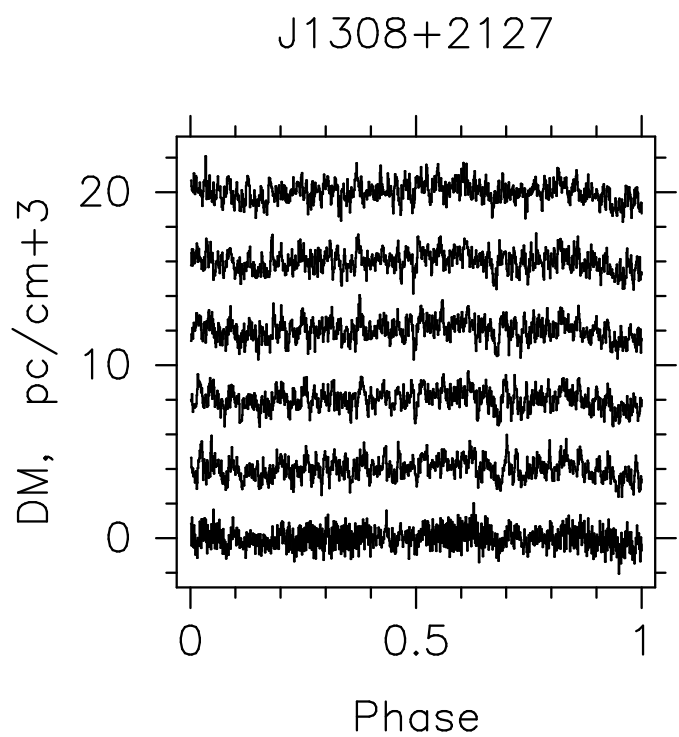}
\caption{
The average profiles of RX~J1308.6+2127 at the frequency of
111~MHz for a number of dispersion measure values.
\label{fig3}}
\end{figure}

\end{document}